\documentclass[sigconf, nonacm]{acmart}

\newcommand\vldbdoi{XX.XX/XXX.XX}

\usepackage{url}
\usepackage{booktabs}

\usepackage{subfigure}
\usepackage{mathtools}
\usepackage{listings}

\usepackage{xcolor}

\definecolor{codegreen}{rgb}{0,0.6,0}
\definecolor{codegray}{rgb}{0.5,0.5,0.5}
\definecolor{codepurple}{rgb}{0.58,0,0.82}
\definecolor{backcolour}{rgb}{0.95,0.95,0.92}

\lstdefinestyle{mystyle}{
    backgroundcolor=\color{backcolour},   
    commentstyle=\color{codegreen},
    keywordstyle=\color{magenta},
    numberstyle=\tiny\color{codegray},
    stringstyle=\color{codepurple},
    basicstyle=\ttfamily\footnotesize,
    breakatwhitespace=false,         
    breaklines=true,                 
    captionpos=b,                    
    keepspaces=true,                 
    numbers=left,                    
    numbersep=5pt,                  
    showspaces=false,                
    showstringspaces=false,
    showtabs=false,                  
    tabsize=2
}

\newcommand{\figwidth}{86mm}
\newcommand{\figspace}{\vspace{0mm}}

\begin{document}
\title{XLPN: Efficient and Scalable Cross-Ledger Protocols for the Topological Consortium of Permissioned Blockchains}

\author{Dongfang Zhao}
\affiliation{%
  \institution{University of Nevada, Reno, United States}
}
\email{dzhao@unr.edu}

\begin{abstract}
While increasingly more application-specific blockchains, or ledgers, are being implemented and deployed,
exchanging information between these ledgers remains an open problem.
Existing cross-ledger protocols (XLPs) exhibit a variety of limitations such as scalability, liveness, efficiency, among others.
This paper proposes a new XLP, namely XLPN-22, 
which introduces a global topology for the consortium of ledgers to achieve better efficiency and scalability of cross-ledger data exchanges.
In this work, we prove the safety and liveness of XLPN-22 and analyze its theoretical complexity.
We also implement XLPN-22 on SciChain ledgers and evaluate it on up to 128 nodes, 8 ledgers, and 16,000 transactions.
Experimental results show that XLPN-22 outperforms two baseline protocols, namely VLDB-20 and PODC-18, by 18--50\% and 64--84\%, respectively. 
\end{abstract}

\maketitle



\section{Introduction}

\subsection{Motivation}

While increasingly more applications (e.g., Internet of Things~\cite{hshen_ndss20}, scientific computing~\cite{aalmamun_sc21}) are adopting permissioned blockchains, or ledgers, as part of their data infrastructure,
the technical challenge of exchanging information between distinct blockchains arised~\cite{dzhao_cidr20}.
The key challenge of such cross-ledger data exchange lies in not only the data itself (which has been actively studied by the database community for decades) but also the consensus protocols employed by each of the ledgers that participate in the information exchange.
To make it worse, the data exchange can happen over an arbitrary number $n$ of heterogeneous ledgers,
essentially forming an $n$-party distributed transaction where each party represents a distributed system, i.e., a permissioned blockchain.
Without an efficient and scalable cross-ledger protocol (XLP) to support the distributed transactions touching on $n$ permissioned blockchains,
application-specific ledgers would be restricted to a small set of functionalities and hardly establish the highly desired consortium of heterogeneous blockchains.

\subsection{Limitations of Existing Approaches}

The exchange between two distinct types of digital currencies (e.g., Ethereum~\cite{ethereum} and Bitcoin~\cite{bitcoin}) has been well supported by the underlying public or permissionless blockchain infrastructure and services~\cite{cosmos}.
However, these techniques are ad hoc for two-party transactions and cannot naturally be extended to an arbitrary number of parties.
Moreover, the latency of such cryptocurrency exchanges is in the order of hours,
which might be fine for financial transactions but is not practical for many applications such as scientific computing applications~\cite{aalmamun_bigdata18}.

Cross-ledger protocols (XLPs) among an arbitrary number of blockchains have also been studied by both the distributed computing and database communities.
Notably, Herlihy et al. (PODC-18~\cite{mherlihy_podc18}) proposed an XLP by traversing the list of ledgers on a ring topology.
However, the ring topology implies that the PODC-18 protocol takes time linear in the number of ledgers,
which is not scalable and will be impractical for a large number of ledgers.

To address the limited scalability among other concerns raised in PODC-18,
Zakhary et al. (VLDB-20~\cite{vzakhary_vldb20}) proposed a two-phase commit (2PC) protocol that views each ledger as a conventional ``node''.
The 2PC coordinator's analog in VLDB-20 is called a \textit{witness chain}.
There is no co-design of intra- and inter-ledger protocols for performance optimization,
and it is unclear how VLDB-20 handles the blocking scenario that is a well-known issue in 2PC.

\subsection{Contributions}

This paper proposes a new XLP, namely XLPN-22, to overcome the limitations of existing protocols as discussed in the previous subsection.
XLPN-22 introduces a global topology for the consortium of ledgers,
which breaks the node hierarchy within and among ledgers such that the number of communication rounds can be reduced from a global point of view.
We prove the safety and liveness of XLPN-22
and analyze its theoretical complexity in terms of both synchronous rounds and the number of messages.
We implement the proposed XLPN-22 protocol as well as the VLDB-20 and PODC-18 protocols on SciChain~\cite{aalmamun_icde21} ledgers,
and evaluate these three protocols on up to 128 nodes, 8 ledgers, and 16,000 transactions.
Experimental results show that XLPN-22 outperforms VLDB-20 by 18--50\% and outperforms PODC-18 by 64--84\%, respectively. 

\section{Background and Related Work}

\subsection{Elementary Topology}

In this paper, we will stick to the terminology and language used in standard topology.
This section gives a quick review and a handy reference for later sections.
We will only discuss the very minimal set of concepts and tools of topology without touching its deep theory and broad applications,
such as homology groups and cohomology rings on manifolds.
There are excellent mathematical texts for this subject, e.g., Munkres~\cite{jmunkres_book84}.

A \textit{topology}, denoted $\mathcal{T}$, is defined as a subset of the superset of a nonempty set $S$,
such that (i) $\{S, \emptyset\} \subseteq \mathcal{T}$,
(ii) an arbitrary union of elements in $\mathcal{T}$ is in $\mathcal{T}$,
and (iii) a finite intersection of elements in $\mathcal{T}$ is in $\mathcal{T}$.
Let $\sigma \in \mathcal{T}$, and $|\sigma|$ denote the number of elements in $\sigma$.
We define the dimension of $\sigma$ as $|\sigma| - 1$, denoted by $dim\;\sigma$.
A \textit{simplex} is a set whose topology includes all the nonempty subsets of its superset.
A $n$-simplex is a simplex whose dimension is $n$;
by definition, this implies that the set underlying the $n$-simplex has $n+1$ elements, each of which is also called a $vertex$.
Any subset of a simplex $\sigma$ is called a \textit{face} of $\sigma$;
in particular, if a vertex $v \in \sigma$ and $\sigma - \{v\} = \tau$, then $\tau$ is called the face \textit{opposite} $v$ in $\sigma$.
A \textit{proper face} of $\sigma$ is a face whose dimension is strictly less than $dim\; \sigma$.
There are two (classes of equivalent) \textit{orientations} associated with a simplex $\sigma = [v_0, v_1, \dots, v_n]$\footnote{Note that we use $[\dots]$ to denote the ordered elements in set $\{\dots\}$.}:
the orientation includes even permutations of $\sigma$ ($\sigma$ is a trivial one since it has 0 permutations of itself),
and the other includes the remainder, i.e., those odd permutations of $\sigma$.
A \textit{complex} is a union of simplices (the plural form of simplex).
The $k$-th skeleton of a complex $\mathcal{C}$, denoted $skel^k(\mathcal{C})$,
is the union of faces whose dimensions are less than or equal to $k$. 
An $n$-simplex is of course a complex as well, usually denoted $\Delta^n$.
The dimension of a complex is defined as that of its highest-dimensional simplex.

We denote $\textbf{D}^n(v)$ the $n$-dimensional \textit{unit disk} centered at vertex $v$, i.e., the set of vertices whose distances to $v$ are less than or equal to 1.
The distance 1 is not necessarily Euclidean distance, any metric space will suffice.
In a Euclidean space $\mathbb{E}^n$, a $\textbf{D}^n$ is simply a solid radius-1 ball at the origin $\textbf{0}^n$.
It is a well-known (and easy to verify) fact that any $\Delta^n$ can be continuously mapped to $\textbf{D}^n$,
and the map is called a \textit{homeomorphism}.
The outer skin, or more formally known the \textit{boundary} $\partial$, of $\textbf{D}^n$ is usually referred to $\textbf{S}^{n-1}$, an $(n-1)$-dimensional \textit{sphere}, or simply an $(n-1)$-sphere.

\subsection{Cross-Ledger Protocols}

In industry, production systems have been developed for exchanging cryptocurrencies.
One such example is Cosmos~\cite{cosmos}, which allows for Bitcoin~\cite{bitcoin} and Ethereum~\cite{ethereum}.
Although Cosmos is built upon an open cross-blockchain protocol named \textit{sidechain}~\cite{sidechain},
no practical systems exist yet for exchanging assets between arbitrary blockchains other than cryptocurrency.
Even for Cosmos and sidechain, criticisms have been widely received regarding the long latency: a cross-blockchain transaction usually takes hours, if not days~\cite{sidechain}, to complete.

For non-cryptocurrency blockchains,
they are usually deployed to a more controlled environment.
Since users of these blockchains need to be screened before being allowed to join the network,
such blockchains are usually referred to as \textit{permissioned blockchains}.
A number of leading service providers such as IBM, Oracle, Azure Blockchain Services, and SAP have made a firm commitment to solving many of the technical challenges that currently plague the interoperability of permissioned blockchains. 
For example, the World Health Organization, in conjunction with the help of the aforementioned companies, was deployed a platform called MiPasa~\cite{mipasa_url20}, which was built atop Hyperledger Fabric~\cite{hyperledger}, to enable ``early detection of COVID-19 carriers and infection hotspots.''~\cite{bcinter_url2020}

\begin{figure}[!t]
    \centering
    \includegraphics[width=\figwidth]{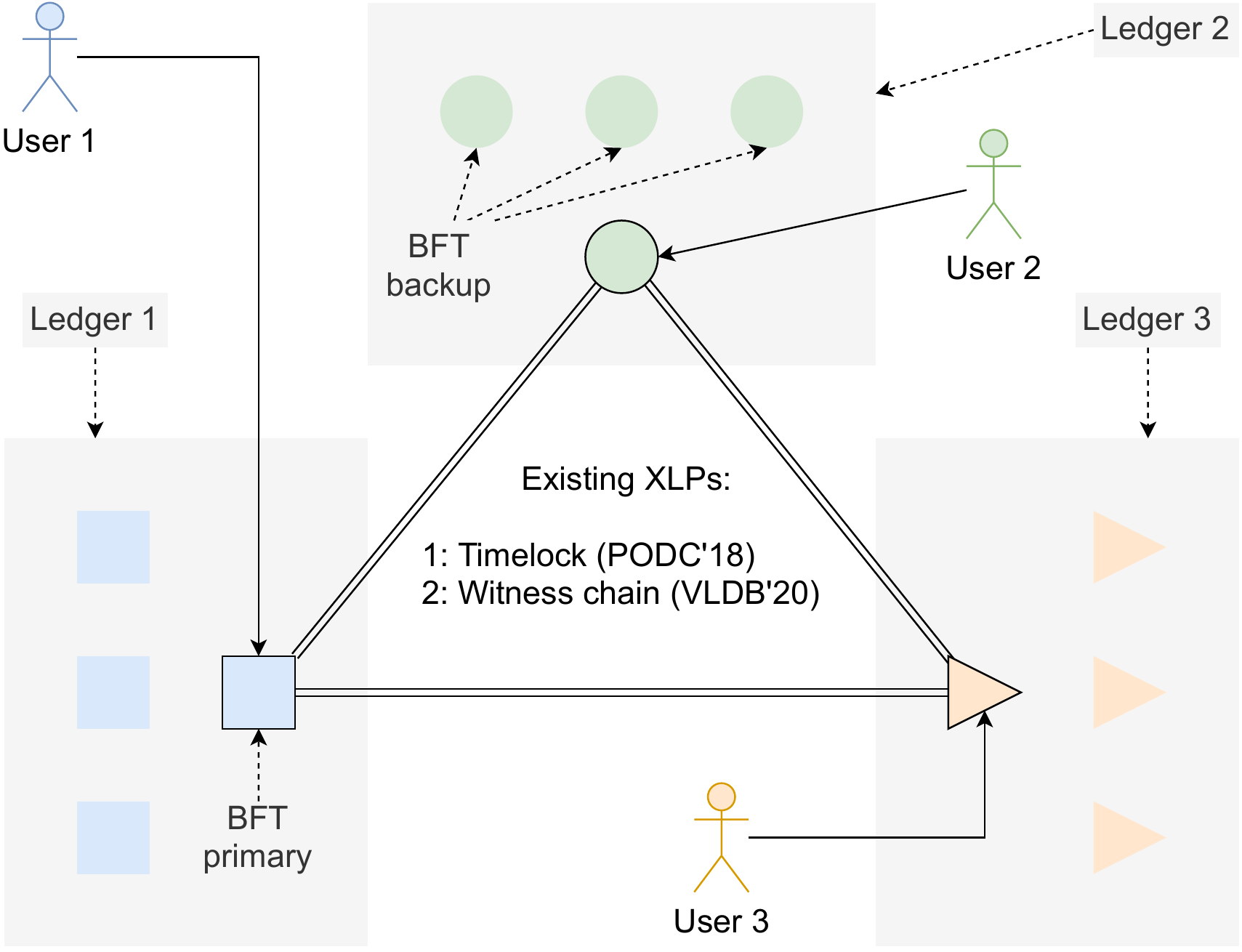}
    \caption{Topological view of existing cross-ledger protocols among three ledgers.}
    \label{fig:xlp_baseline}
     \figspace
\end{figure}

In academia, researchers have been focusing on the cross-ledger protocols (XLPs) among an arbitrary number (i.e., more than two) of permissioned blockchains,
which exhibit a variety of applications such as the Internet of Things (IoT)~\cite{hshen_ndss20} and scientific computing~\cite{aalmamun_sc21}.
At the writing of this paper, two of the most notable protocols were published at PODC-18~\cite{mherlihy_podc18} and VLDB-20~\cite{vzakhary_vldb20}, respectively.
In PODC-18, Herlihy et al. proposed to process the cross-ledger transactions through a timestamp-based approach.
The key idea is to introduce a timeout mechanism, known as \textit{time lock} for the asset to be on hold until the recipient can provide a proof that it qualifies to receive the asset within a predefined period of time.
The approach was then criticized on \textit{atomicity} and \textit{scalability}:
the timeout approach might render some of the parties ``worse off''--- an honest party who sends out its asset and cannot receive compensation due to the network delay (i.e., timeout);
moreover, the timelock requires a sequence of linked smart contracts, leading to a time complexity proportional to the number of parties involved in the transaction.
To address the atomicity and scalability challenges of the above approach, Zakhary et al. in VLDB-20~\cite{vzakhary_vldb20} proposed approaches based on the conventional two-phase commit (2PC) protocol.
The VLDB-20 protocol treats each ledger as a coherent node in its own right and applies 2PC to these nodes by assigning a whiteness chain as the coordinator.
Nevertheless, it is unclear how to overcome the blocking scenario exhibited by 2PC.
Both PODC-18 and VLDB-20 protocols remain theoretical and there is no implementation as of 2021.
We illustrate both protocols in Fig.~\ref{fig:xlp_baseline} in a simple example of 3 ledgers,
each of which reaches a consensus through a Byzantine fault-tolerant (BFT) protocol.

\section{The XLPN-22 Protocol}

\subsection{Description}
\label{sec:protocol_desc}

We formalize the proposed protocol, namely XLPN-22, in this section.
The three communication primitives include inter-ledger one-to-all, inter-ledger all-to-one, and intra-ledger all-to-all.
An inter-ledger one-to-all primitive implies that a single primary server broadcasts a message to each of the primary and backup servers in the ledger cluster.
Similarly, an inter-ledger all-to-one primitive collects a message from all primary and backup servers at a specific primary server.
An intra-ledger all-to-all primitive triggers concurrent executions on all ledgers, 
each execution on a specific ledger involves broadcast messages from each server in that ledger. 

XLPN-22 has five phases of message passing,
not counting the request from the client to the cluster of ledgers.
The first phase is \textit{VOTE-REQ},
which is initiated by the primary of an arbitrary ledger.
The second phase is \textit{VOTE-PREP},
where each server sends and receives a message to and from every server within the same ledger.
The third phase is called \textit{READY},
representing a collection call from all nodes to the initiator of XLPN-22.
The fourth phase is \textit{COMMIT-REQ},
which broadcasts the decision from the initiator to all nodes in the ledger cluster.
The fifth phase is \textit{COMMIT},
where all nodes receive the initiator's decision and reach a consensus to complete the transaction.

\begin{figure}[!t]
    \centering
    \includegraphics[width=\figwidth]{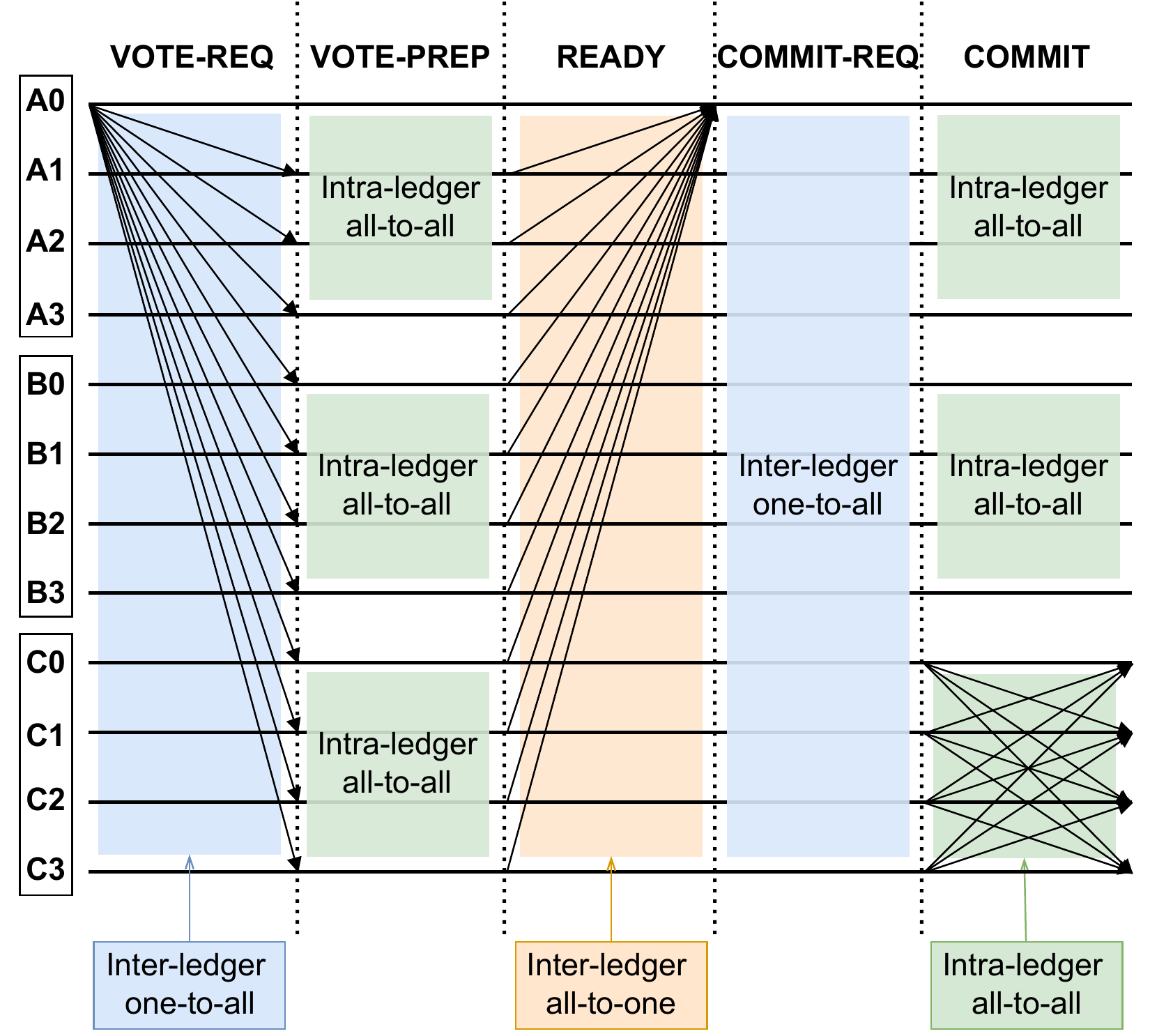}
    \caption{Cross-ledger protocols among three ledgers, each of which comprises four nodes.}
    \label{fig:xlp_topo}
     \figspace
\end{figure}

We illustrate the three communication primitives and five phases in Fig.~\ref{fig:xlp_topo}, 
where XLPN-22 is executed in a cluster of 3 ledgers (i.e., $A$, $B$, and $C$),
each ledger comprising 4 nodes (e.g., $A0$, $A1$, $A2$, and $A3$ for $A$).
For brevity, we only detail the messages once for each of the three communication primitives,
and indicate other instances with rectangle boxes of the same color as that of the detailed messages.
We will elaborate on the five phases of XLPN-22 in the following subsections and refer back to this three-ledger example if needed.

\subsubsection{Phase I: VOTE-REQ}

The transaction requests are collected at a specific node, say the primary server $A0$.
If those requests are submitted at different nodes, it is trivial to aggregate those requests at the designated server.
In the following discussion, we assume the server $A0$ has all the information of the requested transaction spanning over three ledgers $A$, $B$, and $C$.

Unlike the topological model assumed by VLDB-20~\cite{lrup_vldb20} and PODC-18~\cite{pkuz_podc18} that only allows inter-ledger communication among the primary servers,
XLPN-22 allows direct communication cross distinct ledgers.
That is to say, $A0$ can talk to not only 
\{$A1$, $A2$, $A3$, $B0$, $C0$\}
but also \{$B1$, $B2$, $B3$, $C1$, $C2$, $C3$\} as well.
Technically, this can be implemented using either the inter-communicator or the global communicator that are part of the message-passing interface (MPI) standard~\cite{mpi_standard} (more details to be provided in~\S\ref{sec:eval_impl}).

From a topological point of view,
the inter-ledger messages imply that a 1-simplex is added to every pair of vertices residing on distinct ledgers in the complex induced by the ledger cluster.
In fact, because a broadcast message is considered as a communication primitive,
the nodes from all ledgers can be viewed as in the same (subset of) set of the vertices in the complex,
where each vertex represents a specific node in the ledger cluster and the complex represents the cohesion of the ledger cluster.
Let $k$ denote the number of ledgers and $n$ denote the (constant, for simplicity) number of nodes in each ledger.
Those inter-ledger messages essentially translate the ledger cluster into a $(kn-2)$-simplex, or $\Delta^{kn-2}$.
Geometrically, the ledger cluster turns to be equal to $\textbf{S}^{kn-2}$, 
i.e., the boundary of a $(kn-1)$-dimensional disk in the Euclidean space (up to homeomorphism).

\subsubsection{Phase II: VOTE-PREP}

In this phase, all nodes receive the request and prepare for the possible commit of the transaction.
To ensure the local consensus among the nodes within the same ledger,
an all-to-all communication primitive is triggered.
This is illustrated in the second column named VOTE-PREP in the three-ledger example of Fig.~\ref{fig:xlp_topo}.
It should be noted that in conventional Byzantine fault-tolerant (BFT) protocols,
these votes (within each ledger) are communicated in two rounds: PRE-PREPARE and PREPARE.
In XLPN-22, we do not need the PRE-PREPARE round because the VOTE-REQ phase disperses the initial request to all backup servers.

As for topological properties of VOTE-PREP,
because of the all-to-all communication primitive,
each ledger can be viewed as a series of $(n-1)$-simplex,
up to $\lfloor \frac{n-1}{3} \rfloor$ vertices of which can be assigned with a value of either COMMIT or ROLLBACK (due to the Byzantine nodes).
The entire ledger complex has $k$ such simplices,
all of which are disconnected.
Eventually, all nodes in the same ledger will reach a consensus,
leading to a single $(n-1)$-simplex whose vertices are assigned with the same value (e.g., COMMIT) at the end of VOTE-PREP.
For more details of reaching a single simplex in a consensus problem,
we recommend the excellent text~\cite{mherlihy_book13}.

\subsubsection{Phase III: READY}

In this phase, each node replies back to the initiator with its own decision that is agreed upon by at least $2f+1$ nodes (including itself) in the same ledger, 
where $f$ denotes the number of faulty (Byzantine) nodes and $f < \lfloor \frac{n-1}{3} \rfloor$.
From a communication point of view, this phase is basically the reverse of VOTE-REQ.
This phase can be also understood as the analogy of the response phase of the conventional two-phase commit (2PC) protocol.
Indeed, the initiator (e.g., $A0$ in Fig.~\ref{fig:xlp_topo}) works as a coordinator in a 2PC protocol,
and this initiator will make the final call about whether or not to commit the transaction depending on the votes collected from the nodes.
The initiator will decide to commit the transaction only if all nodes agree to commit it;
otherwise, the transaction will be canceled.

Speaking of topology, the all-to-one communication induces another $\Delta^{kn-2}$ complex.
However, if we take into account the direction of the messages,
their orientations are all changed;
for example, the oriented 1-simplex $[A0, A1]$ from phase VOTE-REQ now changes to $[A1, A0]$ in phase READY, which is a 1-permutation of $[A0, A1]$.
There are far-reaching implications of simplex orientation,
notably the Alexander duality,
though we will not discuss this subject any further in this paper.

\subsubsection{Phase IV: COMMIT-REQ}

This phase is yet another one-to-all inter-ledger communication from the initiator to all other nodes.
The purpose of this broadcast is to inform all nodes of the final decision of the transaction, commit or rollback.
This is analogous to the global commit/rollback message from the coordinator to participants in the context of the conventional 2PC protocol.
We skip the topological discussion of this phase as it is identical to the complex discussed previously in phase VOTE-REQ.

\subsubsection{Phase V: COMMIT}

In this last phase, every honest node forwards its received decision from the initiator to other nodes in the same ledger and then collects at least $2f+1$ copies (including that of itself) of the same result.
The honest node finally marks the transaction as completed with the status that is agreed upon by at least $2f+1$ nodes.
It is the client's job (not shown in Fig.~\ref{fig:xlp_topo}) to collect enough information (e.g., $f+1$ copies where $f$ denotes the maximal number of faculty nodes) to be convinced that the transaction is completed.
The topology of this phase is identical to the one of phase VOTE-PREP,
and we skip the details here.

\subsection{Correctness}

The correctness of XLPN-22 relies on two factors:
the intra-ledger phases (VOTE-PREP and COMMIT-REQ) and the inter-ledger phases (VOTE-REQ, READY, and COMMIT).
For each ledger, one implicit requirement is to ensure that the maximal number of faulty nodes is strictly less than one-third of the total nodes.
For simplicity, we assume $n = 3f + 1$.
We also assume a synchronous model because otherwise, consensus cannot be achieved~\cite{mherlihy_stoc93}.

\subsubsection{Intra-ledger phases}

There are three properties that need to be followed in order to reach a consensus within each ledger: agreement, termination, and validity.
We will elaborate on each of them in the following.

\textbf{Agreement}: All honest nodes must agree upon the same value.
This property can be proven in a similar way as that of the PBFT~\cite{mcastro_osdi99} if we assume all messages in XLPN-22 messages are signed by the sending nodes, e.g., ECDSA~\cite{ecdsa}.
Employing the digital signature essentially rules out the possibility for the Byzantine nodes from sending out discrepant messages, e.g., COMMIT for $B0$ and ROLLBACK for $B1$,
because such inconsistency can be immediately detected by multiple receivers (honest nodes) in the ledger.
Recall that in both VOTE-PREP and COMMIT-REQ phases, 
(honest) nodes acquire at least $2f+1$ copies of the same decision.
If there are two different values, say COMMIT and ROLLBACK, that have been agreed by $2f+1$ nodes,
then it implies that $2\times(2f+1) - (3f+1) = f + 1$ nodes have broadcast both COMMIT and ROLLBACK messages.
But this contradicts our assumption that up to $f$ nodes are faulty.
Therefore, both the VOTE-PREP and the COMMIT-REQ phases will reach an agreement.

\textbf{Termination}: All honest nodes must terminate in finite time.
This is trivial to prove:
the intra-ledger all-to-all messages take a single round in VOTE-PREP and COMMIT-REQ,
both of which are in finite time under the assumption of a synchronous model.

\textbf{Validity}: The value that is agreed upon by all nodes must be one of the input values.
In XLPN-22, all the input values are the same because they are either broadcast from VOTE-REQ or COMMIT-REQ.
Therefore, the validity property degenerates into the so-called \textit{all-same validity},
which mandates that the final decision must be the same as the input value.
Then the question becomes whether it is possible for the initial value being COMMIT (ROLLBACK) and the final decision being ROLLBACK (COMMIT).
Let's assume the initial value is COMMIT and the final decision is ROLLBACK for any of the intra-ledger all-to-all primitive.
If the final decision is ROLLBACK, it implies that there are at least $2f+1$ nodes who broadcast ROLLBACK.
Since the protocol ensures that all of $3f+1$ nodes receive the COMMIT input value,
this means that $2f+1$ nodes are faulty,
which contradicts our assumption of up to $f$ faulty nodes.
Therefore, the final decision must be the same as the input value.

\subsubsection{Inter-ledger phases}

Three are two properties that we need to ensure for the three inter-ledger phases, namely the safety and the liveness~\cite{balpern_ipl85}.
Safety says that either the transaction will be committed or aborted in its entirety (atomicity),
and liveness says that the above result will be reached in finite time.
We discuss each perspectively in the following.

\textbf{Safety}: 
All honest nodes will either commit or rollback the transaction.
In the previous subsection, we show that the \textit{agreement} property is guaranteed within each ledger;
therefore, we only need to show that the final decision among different ledgers is consistent.
Without loss of generality, let's assume ledger $B$ decides to commit the transaction while ledger $C$ decides to rollback the transaction, 
in the example execution of Fig.~\ref{fig:xlp_topo}.
According to the \textit{validity} property, this means that some nodes receive a COMMIT command from $A0$ while others receive a ROLLBACK command from $A0$. 
But this would make $A0$ to be considered as a failed node because its messages are digitally signed,
which will trigger a re-election of the initiator, e.g., $B0$.
Therefore, $A0$ could not have sent discrepant values, a contradiction to our assumption.
This proves that the values agreed upon by distinct ledgers are the same.

\textbf{Liveness}: All honest nodes will reach the final decision in finite time.
In normal cases where no failure happens,
XLPN-22 obviously finishes in finite time---within five rounds of message passing.
We need to show that the protocol will nonetheless complete within finite rounds when a failure occurs.
There are three cases to be considered:
the failure of the initiator, 
the failure of primary servers,
and the failure of backup servers.
If the initiator fails, we simply reassign another primary server as the new initiator;
in a synchronous model, this takes one round.
If a primary server fails, a view-change will be triggered,
which takes two rounds (see~\cite{mcastro_osdi99} for details).
If a backup server fails, we take no action because the intra-ledger consensus protocol can tolerate $f$ such failures.
Therefore, the overall number of rounds to handle failures is three,
finite as required.

\subsection{Complexity}

For a fair comparison, we assume the intra-ledger consensus protocols in XLPN-22, VLDB-20, and PODC-18 are all PBFT~\cite{mcastro_osdi99}.
We are interested in two metrics:
the number of synchronous rounds taken by these three protocols and the overall number of messages incurred by them.
Throughout the following discussion, we denote $n$ the number of nodes in each ledger and $k$ the number of ledgers in the consortium.
In the following analysis, we do not consider the failures for brevity;
this can be easily calculated by adding a constant number of two rounds since PBFT requires two rounds for updating the primary servers (i.e., view numbers), namely VIEW-CHANGE and NEW-VIEW.
Each PBFT incurs $(n + n^2 + n^2 + n) = 2n^2 + 2n$ messages all together from its four phases.

\subsubsection{Synchronous rounds}

For PODC-18, there are $2k$ rounds for the round trip of message passing for atomic swap between adjacent pairs of ledgers.
During each two-party swap, there is one PBFT execution that takes 4 rounds (PRE-PREPARE, PREPARE, COMMIT, and REPLY) in civil cases.
If there
So, there are totally $2k \times 4$ rounds for PBFT.
Overall, there are $10k$ rounds in PODC-18.

For VLDB-20, there are 4 rounds for standard 2PC among ledgers.
Within each ledger, one PBFT execution is launched in each phase of 2PC,
resulting in $2\times 4$ rounds of message passing.
Therefore, there are overall $4 + 2\times 4 = 12$ rounds of message passing in VLDB-20.

For XLPN-22, there are overall five synchronous rounds (see~\ref{sec:protocol_desc}).
We summarize these results in the middle column of Table~\ref{tbl:comparison}.

\begin{table}[t]
  \begin{center}
    \caption{Complexity comparison of cross-ledger protocols. Suppose there are $k$ ledgers, each of which runs PBFT among its $n$ participating nodes.}
    \label{tbl:comparison}
    \begin{tabular}{l c l}
\toprule
    Protocol                            & \# of rounds    &   \# of messages \\
\midrule
    PODC-18~\cite{pkuz_podc18}          &   10$k$   & $4kn^2 + 4kn + 4$ \\
    VLDB-20~\cite{vzakhary_vldb20}      &   12      & $4kn^2 + 4kn + 4k$ \\
    XLPN-22 (this work)                 &   5       & $4kn^2 + 3kn + 4k - 3$ \\
\bottomrule
    \end{tabular}
  \end{center}
\end{table}

\subsubsection{Number of messages}

For PODC-18,
each two-party swap involves 2 messages plus one PBFT call,
resulting in $2n^2 + 2n + 2$ messages.
Since there are $2k$ such two-party swaps,
the overall number of message is $2k\times (2n^2+2n+2) = 4kn^2 + 4kn + 4$.

For VLDDB-20,
the standard 2PC protocol between ledgers incurs $2\times 2k = 4k$ messages.
Within each ledger, $2k$ PBFT executions are launched,
leading to $2k \times (2n^2 + 2n) = 4kn^2 + 4kn$ messages.
Therefore, the overall messages are $4kn^2 + 4kn + 4k$.

For XLPN-22, two intra-ledger phases incur $2k \times (2n^2 + 2) = 4kn^2 + 4k$ messages.
Three inter-ledger phases incur $3 \times (kn-1) = 3kn -3$ messages.
The overall number of messages is $4kn^2 + 3kn + 4k - 3$. 
We summarize the results in the right-most column of Table~\ref{tbl:comparison}.

\subsection{Recovery}

The failure of the initiator would cause the transaction to rollback,
which can be easily detected by a heartbeat protocol.
Unlike conventional 2PC protocols, we can appoint the primary server of another ledger as the new initiator without waiting for the recovery of the original initiator.
The detail has been discussed in the liveness property in the previous subsection.

The failure of the primary server within each ledger will trigger a view-change of all ledgers.
This is different from conventional BFT protocols because we now need to synchronize the new view number across the consortium of ledgers.
For brevity, we do not explicitly formalize the view-change procedure,
which can be found in any standard text of distributed systems, e.g.,~\cite{dspp}.

\section{Evaluation}

\subsection{Implementation}
\label{sec:eval_impl}

We implement the proposed protocol (XLPN-22), as well as two baseline protocols (VLDB-20~\cite{vzakhary_vldb20} and PODC-18~\cite{mherlihy_podc18}), on a cluster of SciChain instances~\cite{aalmamun_icde21}.
SciChain is a blockchain prototype crafted for scientific data management (especially for provenance),
which is presumably to be deployed as a permissioned distributed ledger.
SciChain was implemented with the Message Passing Interface (MPI) through a Python language binding called mpi4py~\cite{mpi4py},
which is built on Open MPI~\cite{openmpi} that is implemented mostly with C and Shell script.
This work contributes about 1,500 lines of code.

The original SciChain implementation did not allow inter-ledger communication because a SciChain instance is managed by a global communicator in its own right.
When implementing the cross-ledger protocols, we create regional communicators for different groups of nodes,
each of which represents a distinct ledger.
Obviously, those groups collectively form a partition of the set of all participating nodes managed by the global communicator.
By doing so, intra-ledger communication can go through the regional communicator while inter-ledger communication can be carried out by the global communicator.
An alternative implementation could be directly creating multiple groups of processes and using the inter-communicator primitives for inter-ledger communication.
The latter implementation usually provides higher flexibility for specifying the node topology but with some performance overhead.

\subsection{Experimental Setup}

All experiments were carried out on the CloudLab testbed~\cite{cloudlab}.
We use the \texttt{c6420} servers,
each of which is equipped with Intel Xeon Gold 6142 CPUs at 2.6 GHz, 384 GB ECC DDR4-2666 memory, and 	
two Seagate 1~TB 7200 RPM 6G SATA HDDs.
Each server is connected via a 1 Gbps control link (Dell D3048 switches) and a 10 Gbps experimental link (Dell S5048 switches).
The operating system image is Ubuntu 20.04.3 LTS.

We use the ByShard transaction workload~\cite{byshard_vldb21} as our benchmark (i.e., the cross-shard transactions are now considered cross-ledger transactions),
but do not count the execution time other than message passing.
If not otherwise stated, the default number of transactions in the experiments is 5,000.
When evaluating the performance impact of transaction workloads, however,
we vary the number of transactions from 1,000 to 16,000.
By default, there are 4 ledgers, each of which consists of up to 32 nodes.
When evaluating the impact of various numbers of ledgers,
we set 16 as the number of nodes in each ledger.
Overall, we use up to 128 nodes on CloudLab~\cite{cloudlab}.
All experiments are repeated three times, and we report the average numbers along with their standard errors.

\subsection{Varying Numbers of Transactions}

The first experiment reports the protocol performance as a function of varying numbers of transactions.
We fix the ledger environment by processing those transactions on 4 ledgers and each ledger has 32 nodes.
We expect that the overall processing time will be proportional to the number of transactions
unless the network bandwidth is saturated,
which should not happen to the given workload.

\begin{figure*}[t]
\minipage{0.32\textwidth}
    \includegraphics[width=\linewidth]{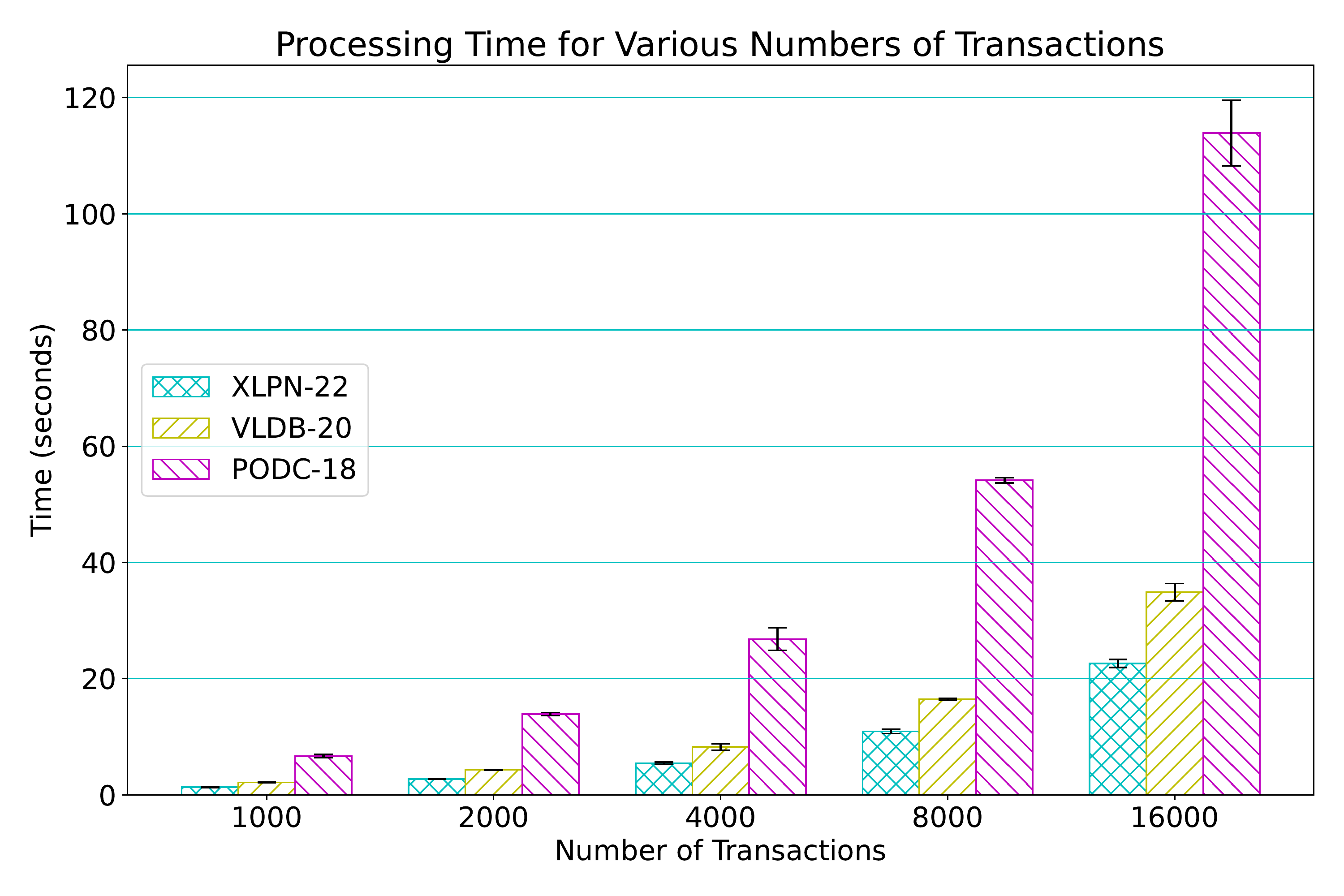}
    \caption{Performance comparison of three protocols when processing various numbers of transactions.
    There are 4 ledgers, each having 32 nodes.}
    \label{fig:txn}
\endminipage\hfill
\minipage{0.32\textwidth}%
    \includegraphics[width=\linewidth]{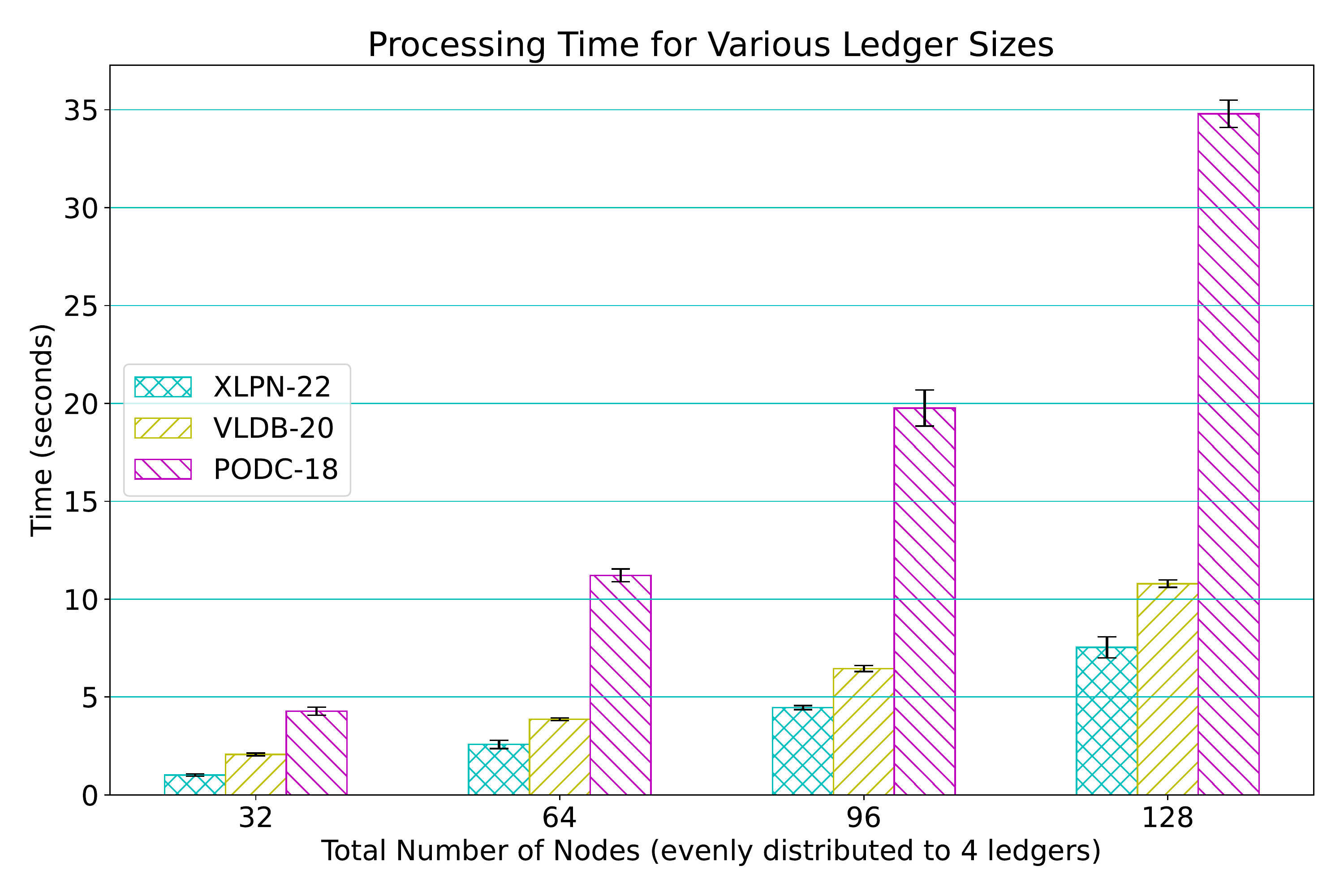}
    \caption{Performance comparison of three protocols when 5,000 transactions are submitted to 4 ledgers,
    each of which has 8, 16, 24, and 32 nodes, respectively.}
    \label{fig:miner}
\endminipage\hfill
\minipage{0.32\textwidth}%
    \includegraphics[width=\linewidth]{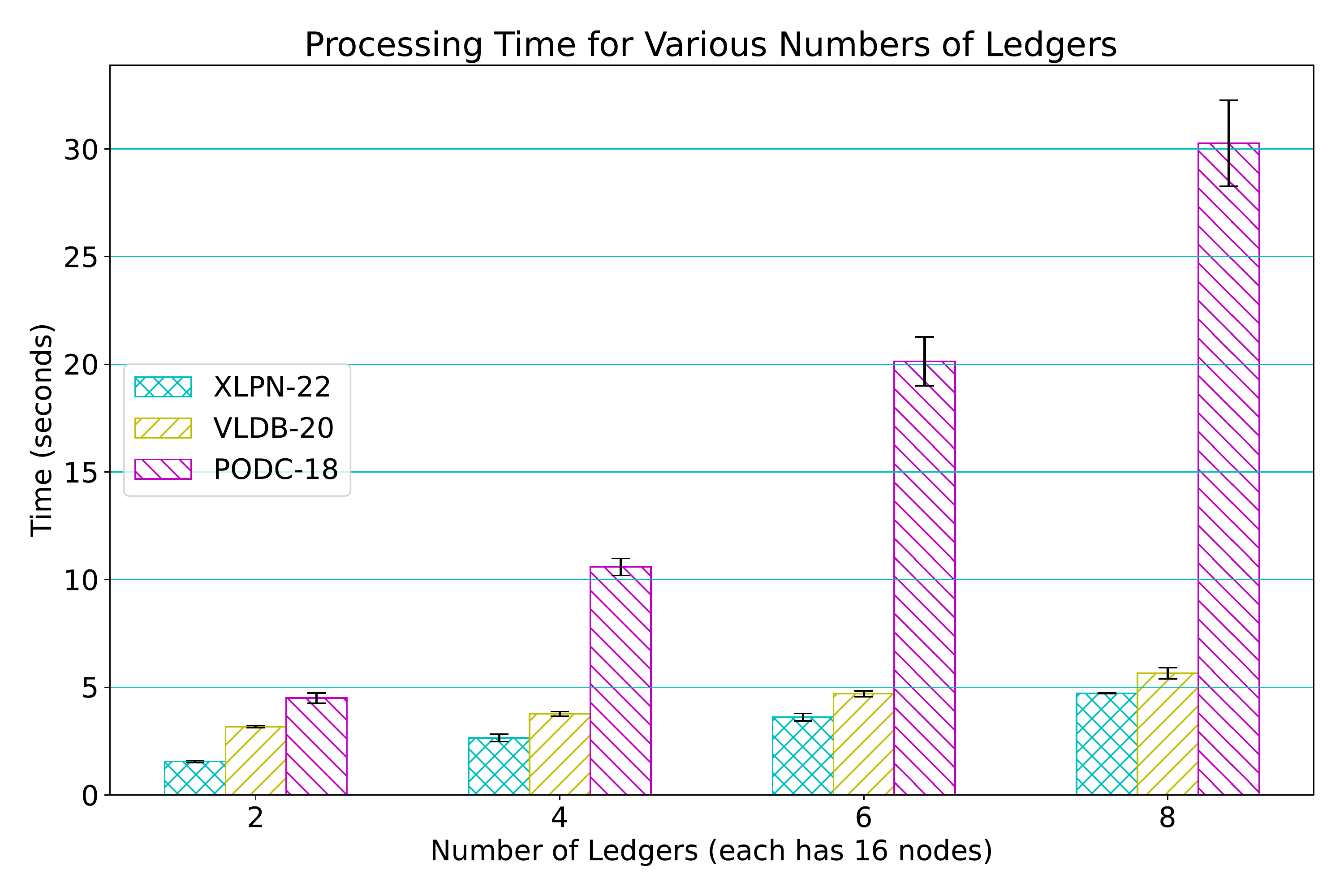}
    \caption{Performance comparison of three protocols when 5,000 transactions are submitted to a consortium of 2, 4, 6, and 8 ledgers; each ledger has 16 nodes.}
    \label{fig:chain}
\endminipage
\end{figure*}


Fig.~\ref{fig:txn} reports the transaction processing time of three protocols.
We observe that all three protocols exhibit an approximately linear increase in processing time when being fed more transactions.
The results suggest that XLPN-22 incurs the shortest processing time.
For example, XLPN-22 takes 22.6 seconds to process 16,000 transactions, while VLDB-20 and PODC-18 take 34.9 and 113.9 seconds to complete the same workload;
this makes the XLPN-22 protocol 35\% and 80\% faster than VLDB-20 and PODC-18, respectively.

\subsection{Varying Numbers of Nodes}

The second experiment investigates the performance impact from the ledger sizes,
which refer to the number of nodes admitted to a specific permissioned ledger.
For this preliminary experiment, we assume that (i) there are 4 ledgers in total and (ii) the number of nodes in each ledger is the same and is selected from one of the following values:
8, 16, 24, and 32.
This is equivalent to saying that the total numbers of participating nodes are 32, 64, 96, and 128, respectively.
Also recall that the default number of transactions is used in this experiment, which is 5,000.

Fig.~\ref{fig:miner} reports the performance of the three protocols at different scales of nodes.
As expected, all protocols show that the processing time increases monotonically with respect to the ledger size:
The more nodes that are involved in a distributed transaction,
the more time that needs to be taken for the transaction to complete.
Unlike the performance impact from numbers of transactions,
the ledger size seems to weigh more:
the constant increment (i.e., 32 nodes) roughly implies a doubled time span of the execution (as opposed to the case that a doubled increment implies a doubled time span).

Unsurprisingly, the proposed XLPN-22 protocol outperforms the baseline protocols by a significant margin (we have seen this when discussing the performance of varying numbers of transactions).
At the largest 128-node scale, for example,
XLPN-22 takes 7.5 seconds to complete 5,000 transactions while VLDB-20 takes 10.8 and PODC-18 takes 24.8 seconds,
which implies that XLPN-22 is 31\% and 70\% faster than VLDB-20 and PODC-18, respectively.

\subsection{Varying Numbers of Ledgers}

This section reports the performance impact made by the different numbers of ledgers to which the distributed transaction is submitted.
To isolate other factors, we fix the total number of transactions as the default one (5,000) and we fix the ledger size as 16---each ledger comprises 16 nodes.
We vary the numbers of ledgers from the trivial case, i.e., two-ledger transactions, to eight distinct ledgers (again, each having 16 nodes).
That is, the largest scale of experiments involves 128 nodes.

Fig.~\ref{fig:chain} reports the performance of the three protocols when 5,000 transactions are submitted to a federation of 2, 4, 6, and 8 distinct ledgers, each of which has 16 participating nodes.
The performance of PODC-18 shows a similar performance pattern as the one for ledger sizes (cf.~Fig.~\ref{fig:miner}):
the impact from the number of ledgers also exhibits the pattern that a constant factor increment (i.e., 2 more ledgers) implies doubled processing time.
However, XLPN-22 and VLDB-20 exhibit a more mild impact than PODC-18.
In fact, the performance gap between XLPN-22 and VLDB-20 on more ledgers is narrowed.
Specifically, for a two-ledger transaction,
XLPN-22 completes 5,000 transactions in 1.6 seconds,
VLDB-20 and PODC-18 complete the same workload in 3.2 and 4.5 seconds;
for an 8-ledger transaction,
XLPN-22 completes 5,000 transactions in 4.7 seconds,
VLDB-20 and PODC-18 complete the same workload in 5.7 and 30.1 seconds.
In other words, XLPN-22 outperforms VLDB-20 and PODC-18 by 50\% and 64\% on two ledgers,
but then the speedup changes to 18\% and 84\%, respectively, on eight ledgers.
This result suggests that for those transactions involving many ledgers (e.g., more than 10),
VLDB-20 may be a more efficient protocol (given the condition that all other factors are equal).
We leave this as an open question and will address it in our future work.

\section{Conclusion and Future Work}

This paper presents a new cross-ledger protocol, namely XLPN-22, for a consortium of permissioned blockchains.
XLPN-22 advances state-of-art protocols by co-designing the intra- and inter-ledger protocols for distributed transactions.
We analyze the topological properties of XLPN-22 and derive its theoretical complexity in terms of both synchronous rounds and the number of messages.
We experimentally verify the effectiveness of XLPN-22 by comparing it against two baseline protocols, VLDB-20 and PODC-18.

This work remains at an infancy stage along this line of research.
Our future work will include:
(i) Conducting more extensive experiments with real-world workloads, more ledgers (e.g., 1,000+), and other metrics such as memory footprint and network utilization, so that we can characterize XLPN-22 regarding its performance bottleneck, applicability, and other properties;
(ii) Employing more topological tools to better understand XLPN-22 such that in the future we could co-design the protocol with its topological features to meet application-specific needs; and
(iii) Investigating the feasibility of applying XLPN-22's inter-ledger phases to permissionless blockchains,
e.g., how to replace XLPN-22's BFT consensus protocols with proof-of-work mechanisms without compromising the correctness and performance of XLPN-22.

\bibliographystyle{acm}
\bibliography{ref_new}

\end{document}